\documentclass[superscriptaddress,twocolumn,aps,prb,floatsfix]{revtex4}
\usepackage{graphicx}
\usepackage{amssymb,amsmath}
\usepackage{array}
\usepackage{dcolumn}
  
\begin{document}
 
\title{An ab-initio evaluation of the local effective interactions in the $\rm
  Na_{x} Co O_2$  family}
 
\author{Sylvain Landron}
\author{Julien Soret}
\author{Marie-Bernadette Lepetit}
 
\affiliation{CRISMAT, ENSICAEN-CNRS UMR6508, 6~bd. Mar\'echal Juin, 14050 Caen, 
FRANCE}

\date{\today}

\begin{abstract}
  We used quantum chemical ab initio methods to determine the effective
  parameters of Hubbard and $t-J$ models for the $\rm Na_{x}CoO_2$ compounds
  (x=0 and 0.5). As for the superconducting compound we found the $a_{1g}$
  cobalt orbitals above the $e_g^\prime$ ones by a few hundreds of meV due ti
  the $e_g^\prime$--$e_g$ hybridization of the cobalt $3d$ orbitals.  The
  correlation strength was found to increase with the sodium content $x$ while
  the in-plane AFM coupling decreases. The less correlated system was found to
  be the pure $CoO_2$, however it is still strongly correlated and very close
  =to the Mott transition. Indeed we found $U/t\sim 15$, which is the critical
  value for the Mott transition in a triangular lattice.  Finally, one finds
  the magnetic exchanges in the $\rm CoO_2$ layers, strongly dependant of
  the weak local structural distortions.
\end{abstract}
\maketitle                                                                            

\section{Introduction} \label{intro}
Since the discovery of the large thermoelectric power~\cite{TEP} and then of
superconductivity~\cite{supra} in the $\rm Na_x Co O_2$ layered cobaltate
family, these systems have attracted a lot of attention. Indeed, they present
a very rich phases diagram, as a function of the sodium content $x$, and the
possible water intercalation.  
For small $x$ ($0<x<0.5$) the magnetic susceptibility does not depend on
temperature. These systems are thus supposed to be paramagnetic metals,
however some authors reinterpreted the data (using NMR
measurements~\cite{DiagPh2}) and conclude that these systems should be
antiferromagnetic metals.  For larger $x$ ($0.5<x<0.75$) the compounds are
Curie-Weiss metals. Over $x=0.75$, the system has be seen as a weakly
ferromagnetic metal~\cite{DiagPh}, however neutron diffraction
measurements~\cite{AFM_A} rather sees A-type antiferromagnetism, that is
ferromagnetic $\rm CoO_2$ layers, antiferromagnetically coupled. 
For special values of the sodium content, $x=1/4$~\cite{IR05_025},
$x=1/2$~\cite{IR05_025} and $x=1$~\cite{Na1CoO2}, the system is insulating. A
band insulator for $x=1$ and a charge/spin ordered state for $x=0.5$. On the
contrary, for $x=0$ the system remains conducting~\cite{CoO2_1,CoO2_2,CoO2_3},
exhibiting a Fermi liquid behavior, while its half-filling character
associated with the large electron correlation of the cobalt $3d$ let expect a
Mott insulator.

The $\rm Na_x Co O_2$ compounds are composed of $\rm Co O_2$ layers in
the $(\vec a, \vec b)$ plane, between which the sodium ions are
located.  The $\rm Co O_2$ layers are formed by $\rm CoO_6$
edge-sharing octahedra forming a two-dimensional triangular lattice of
cobalt ions. The $\rm CoO_6$ octahedra present a trigonal distortion
along the crystallographic $\vec c$ axis corresponding to a
compression of the oxygen planes toward the cobalt one. Additional
distortions take place, according to the sodium ordering. 

In the pure $\rm Co O_2$ compound, although Tarrascon {\it et
  al}~\cite{Tarrascon} suggested a complex crystallographic structure, further
authors refined the structure with a single cobalt crystallographic
position~\cite{CoO2_1,CoO2_2}. This system exhibits a geometry quite
characteristic compared to the rest of the $\rm Na_x CoO_2$ family. First, the
inter-layer distance is very short. Indeed, the inter-layer distance is
4.48\,\AA\ , 4.31\,\AA\ or 4.25\,\AA\ according to the authors, while for
$x\ne 0$ it ranges from 5.2\,\AA\ to 5.7\,\AA\ and increases  with
decreasing $x$. Second, the $\rm CoO_2$ layer is very compressed
in the $\vec c$ direction. Indeed, its thickness is only 1.71\,\AA\, while it
ranges between 1.92\,\AA\ and 1.98\,\AA\ for the $\rm Na_x Co O_2$ family. In
fact, in this system, the $\rm CoO_2$ layer is even more compressed than in
the superconducting, hydrated compound, for which its thickness is
1.76\,\AA. It results in a very large trigonal distortion of the $\rm CoO_6$
octahedra, with 62.3 degrees between the $\vec c$ axis and the Co--O
directions. Using  the results of reference~\onlinecite{nous2}, this strong
trigonal distortion lets us expect a strong
$a_{1g}$--$e_g^\prime$ splitting. Third, the Co--O distances are shorter than
in all the other compounds of the family, with only 1.847\,\AA.

\begin{figure}[h]
\begin{minipage}{3.5cm}
$\rm \bf Na_{0.5} CoO_2$ \\
\resizebox{3.2cm}{!}{\includegraphics{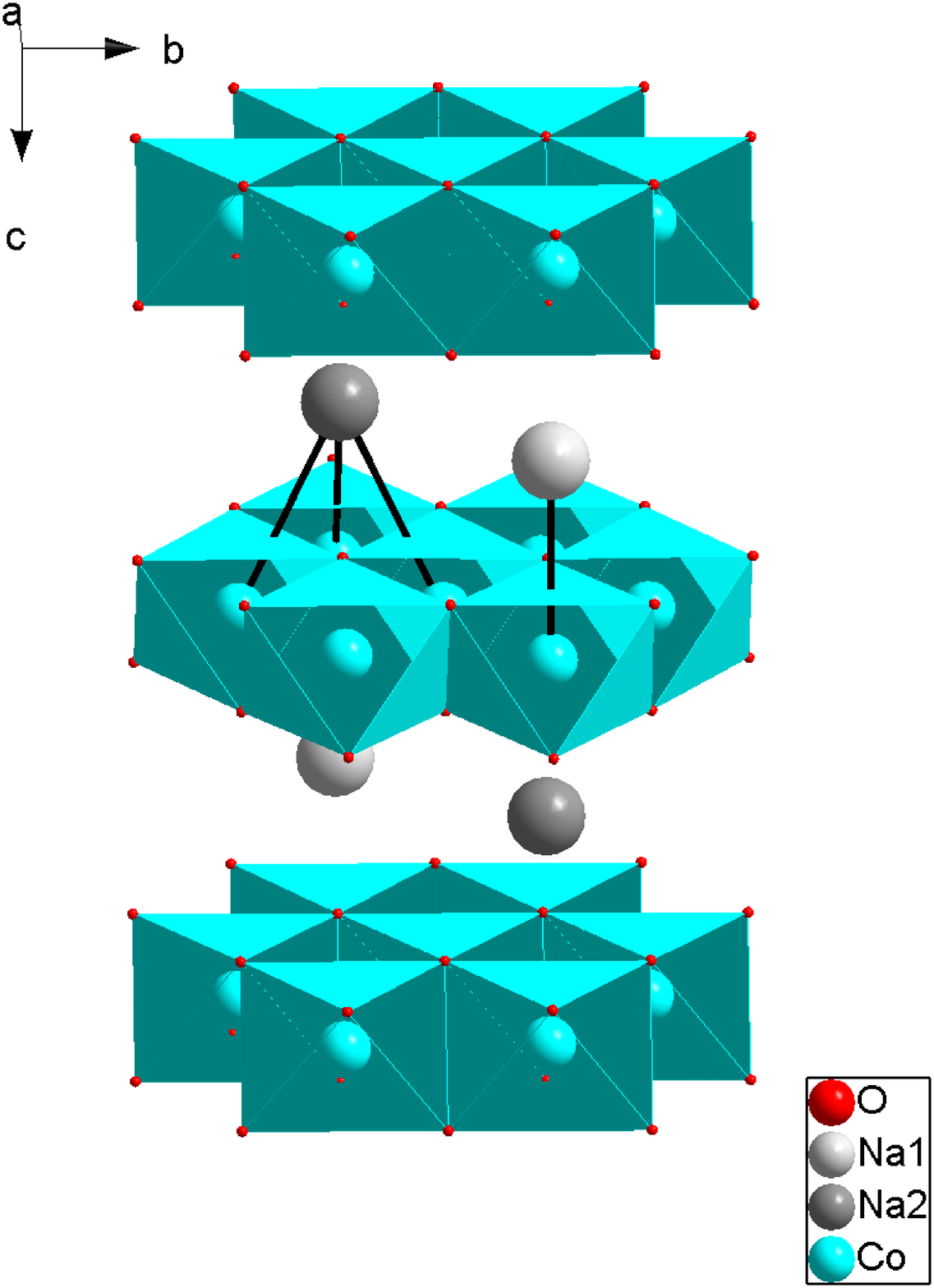}}
\end{minipage}
\begin{minipage}{4.5cm}
$\rm \bf CoO_2$ \\
\resizebox{4cm}{!}{\includegraphics{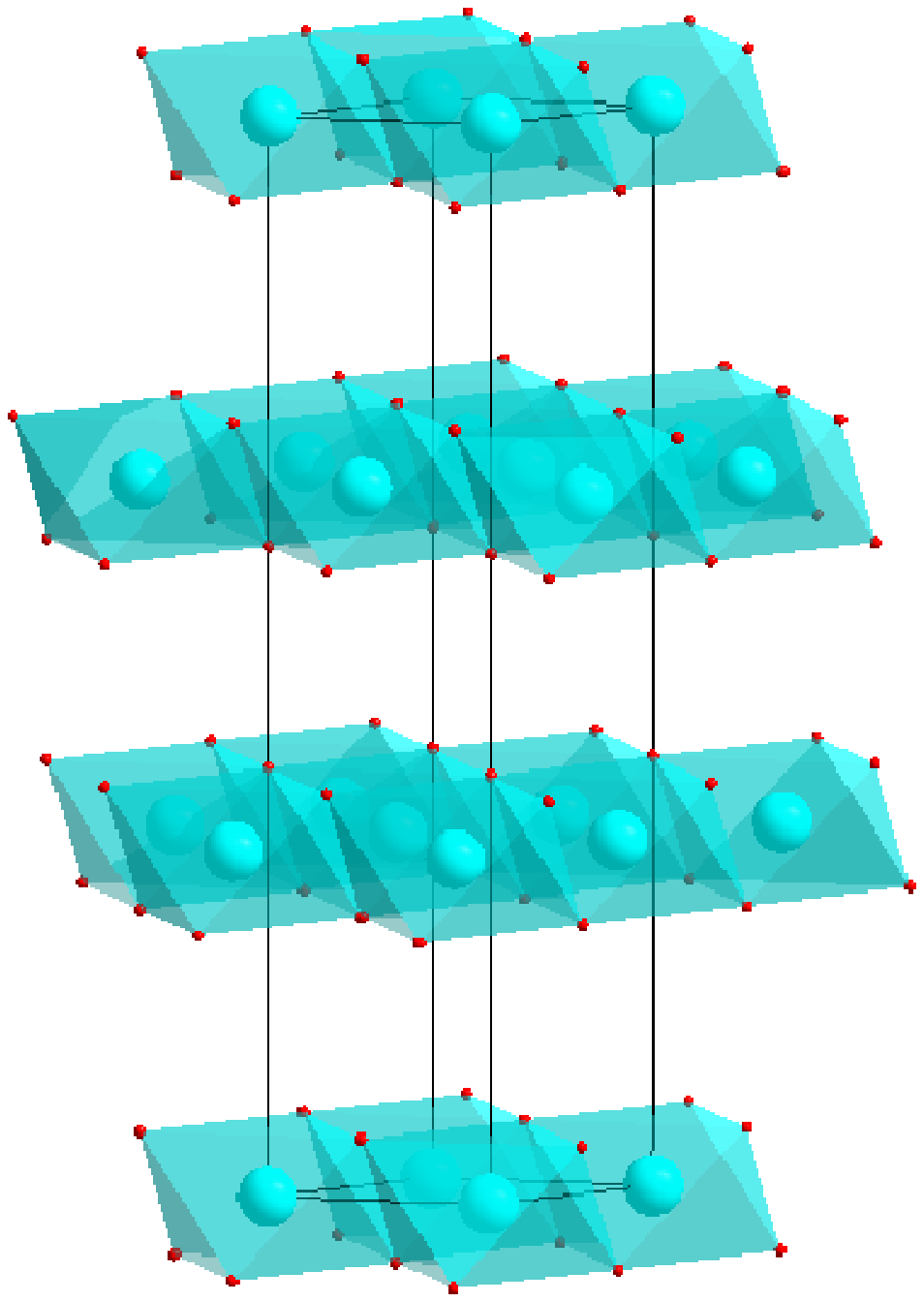}}
\end{minipage}
\caption{Schematic structure of the $\rm Na_{0.5} Co O_2$ and $\rm Na_{0} Co O_2$
compounds. The black lines connect the $\rm Na^+$ ions with their
nearest cobalt neighbors.}
\label{f:struct}
\end{figure}

In the $\rm Na_{0.5} Co O_2$ compound (P2 structure, subject of the present
work), the sodiums ordering is commensurate and presents two types of
alternated stripes along the $\vec b$ direction~\cite{struct05}. In the first
stripes, the sodium atoms are located on-top of (below) the cobalts atoms, and
in the second type of stripes the sodium ions are located on-top of (below)
the center of the cobalt triangles (see figure~\ref{f:struct}).  It results
two types of cobalt atoms organized in stripes along the $\vec b$ direction,
namely the Co(1) atoms (located on top of a sodium ion) and the Co(2)
ones. Compared to the other systems of the family, the $\rm Na_{0.5} Co O_2$
compound does not stick out~\cite{structx}. Indeed, its $\rm CoO_2$ layers
present a thickness of 1.97\,\AA\ . The average Co--O distances are 1.887\AA\
for Co(1) and 1.900\AA\ for Co(2). The average trigonal distortions are
associated with angles of 58.9 degrees (between the $\vec c$ axis and the
Co(1)--O directions), and 59.4 degrees (Co(2)--O). These angles are weaker
than for the superconducting system or the pure $\rm CoO_2$ one, but still
quite larger than the 54.7 degrees of the regular octahedron. One can thus
expect a splitting of the $t_{2g}$ cobalt $3d$ orbitals in two
quasi-degenerate $e_g^\prime$-type orbitals of low energy and a $a_{1g}$-type
one at the Fermi level.

Out of all the $\rm Na_{x} Co O_2$ systems, the $x=0.5$ compound 
presents a very rich but not well understood phases diagram. Indeed, the
$\rm Na^+$ ions order at very large temperature, namely slightly above
300~K~\cite{ordreNa}, inducing a small charge order in the cobalt
layer. Despite this small charge order the system remains
conducting. At 88~K a magnetic phase transition occurs, associated with
a structural distortion, and the onset of a long range magnetic
order. Again, despite this transition, the system remains conducting up
to 53~K where a small charge gap (14 meV) finally
opens~\cite{IR05_025,cascade}. 
The understanding of this complex phase diagram is still under
process. Different hypotheses have been advanced such as successive
Fermi surface nesting~\cite{nesting}, polarons
formation~\cite{polarons}, one band versus three bands
cross-over~\cite{picket}. 

As already mentioned, the $\rm CoO_2$ compound remains conducting for all
temperatures, opposite to the Mott insulator behavior that was expected due to
its half-filled character.  The low temperature dependence of the magnetic
susceptibility lead some authors to suggest that this compound is a weakly
correlated metal~\cite{CoO2_2}. However, as for the other compounds with $x
\leq 0.5$, these data were reinterpreted~\cite{CoO2_1} in view of
NMR measurements as due to strong antiferromagnetic coupling.

In order to understand the low energy physics of these systems it is thus
necessary to have accurate determinations of the pertinent degrees of freedom
and of their coupling, as a function of the exact crystallographic
structure. Indeed, all authors agree on the importance of the sodium ordering
and the induced cobalt local environment distortions, on the low energy
properties. The aim of this paper is to provide such informations using state
of the art quantum chemical spectroscopy methods for the $x=0.5$ and $x=0$
compounds. The next section will shortly recall the methods. Section III will
presents our results and finally the last section will sum up our conclusions.

\section{Method and computational details} \label{methode}
\subsection{Ab initio calculations}\label{meth:abinitio}

The method used in this work (CAS+DDCI~\cite{DDCI}) is a configurations
interaction method, that is an exact diagonalization method within a
selected set of Slater determinants, on embedded crystal fragments.
This method has been specifically designed to accurately treat
strongly correlated systems, for which there is no single-determinant
description. The main point is to treat exactly all correlation
effects and exchange effects within a selected set of orbitals (here
the $3d$ shell of the cobalt atoms) as well as the excitations
responsible for the screening effects on the exchange, repulsion,
hopping, etc.  integrals.

The CAS+DDCI method has proved very efficient to compute, within
experimental accuracy, the local interactions (orbital energies, atomic
excitations, exchange and transfer integrals, coulomb repulsion etc.)
of a large family of strongly correlated systems such as high $T_c$
copper oxides~\cite{DDCIhtc}, vanadium oxides~\cite{vana}, nickel and
cuprate fluorides~\cite{DDCI_Ni}, spin chains and
ladders~\cite{incom}, etc.

\begin{figure}[h] 
\resizebox{!}{2cm}{\includegraphics{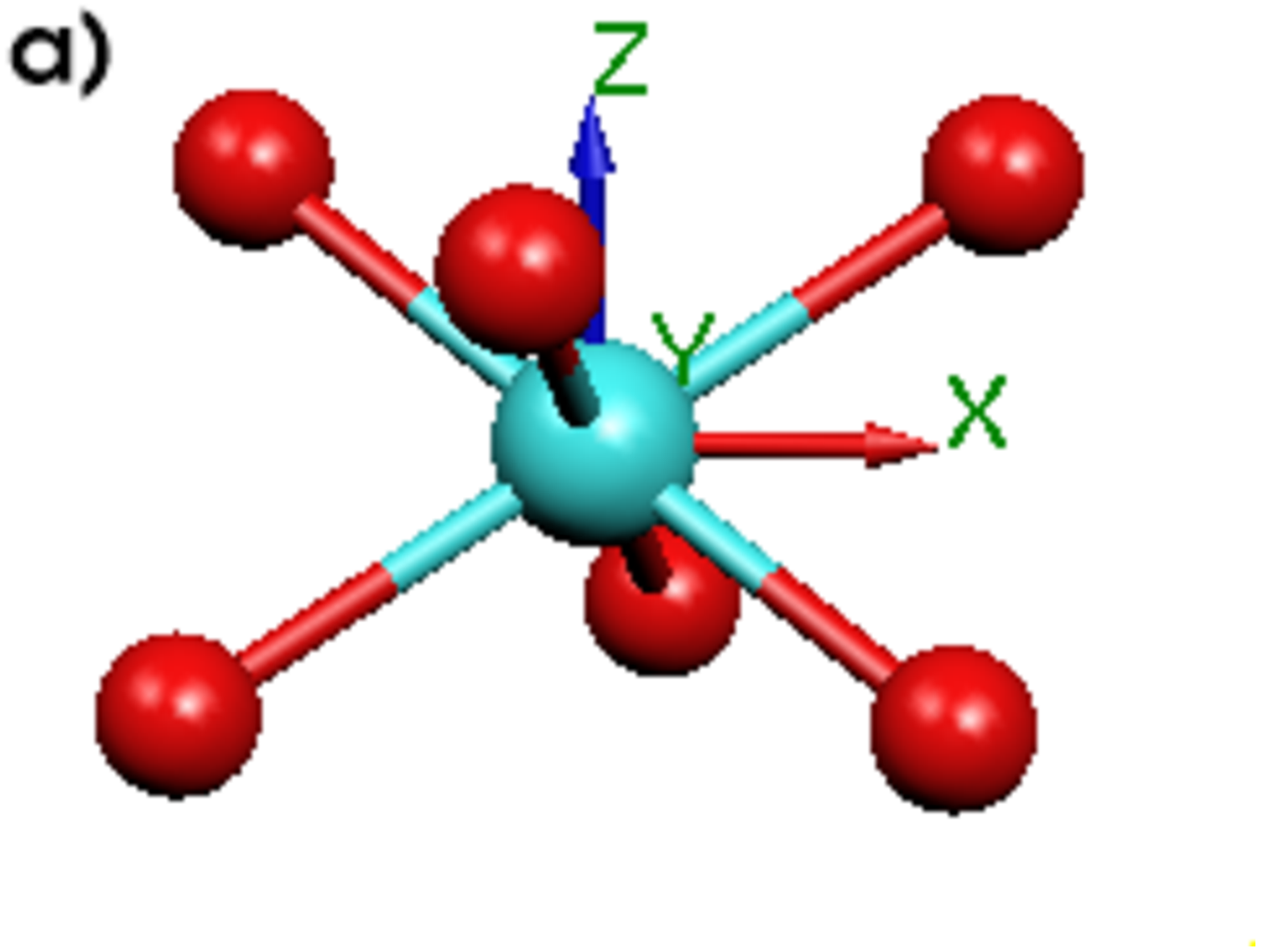}}
\resizebox{!}{2cm}{\includegraphics{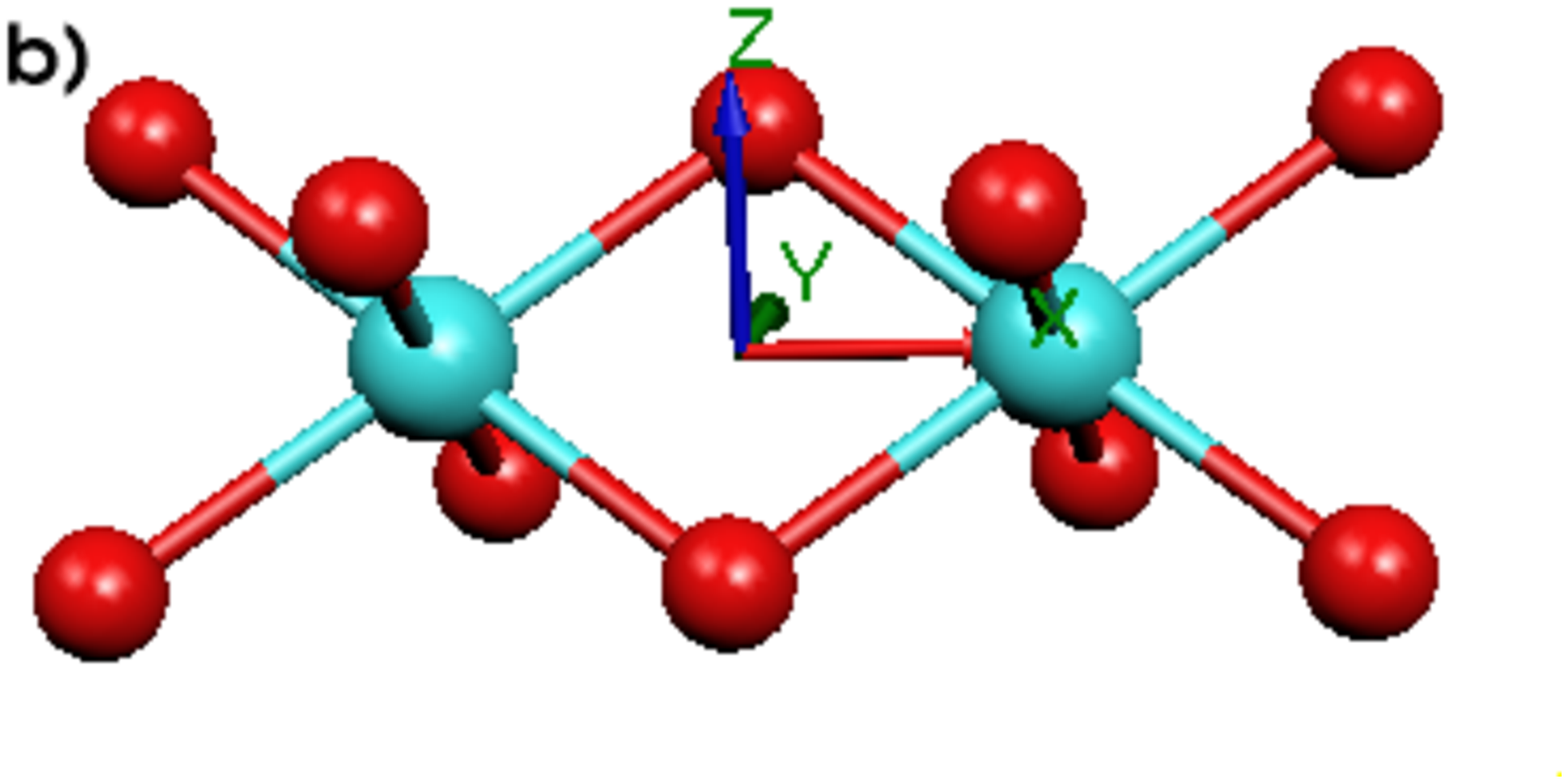}}
\caption{a) $\rm CoO_6$ and b) $\rm Co_2 O_{10}$ fragments used in the
present calculations.}
\label{f:clus}
\end{figure}

The clusters used in this work involve one and two cobalt atoms ($\rm
CoO_6$ and $\rm Co_2O_{10}$) and their oxygen first coordination
shell (see figure~\ref{f:clus}). These fragments are embedded in a
bath designed so that to reproduce on them the main effects of the
rest of the crystal~; that is the Madelung potential and the exclusion
effects of the electrons of the other atoms of the crystal on the
clusters electrons.

The electrostatic potential is reproduced by a set of point charges located at
the atomic positions. The charges are renormalized next to the bath borders,
using the Gell\'e's method, in order to obtain an exponential convergence of
the Madelung potential~\cite{env2}. The convergence accuracy was set in the
present work to the mili-electron-Volt.  The nominal atomic charges used in
this work are the formal charges, that is $+3.5$ and $+4$ for the cobalt atoms
of the $x=0.5$ and $x=0$ compounds, $-2$ for the oxygen atoms and $+1$ for the
sodium atoms.  The exclusion effects are treated using total ions
pseudo-potentials~\cite{TIPS} (TIP) on the first shell of atomic sites
surrounding the clusters.

The calculations were done using the MOLCAS~\cite{molcas} and
CASDI~\cite{casdi} suites of programs. The basis sets used can be found in
reference~\onlinecite{bases}. The structural parameters were taken for the $\rm
Na_{0.5}CoO_2$ system from the 10~K data of reference~\onlinecite{struct05}
and for the $\rm Na_{0}CoO_2$ system from reference~\onlinecite{CoO2_2}.

\subsection{Extracting effective interactions from ab initio results}
\label{meth:param}

The results of the ab initio calculations on an embedded fragment of the
crystal provide us with energies ($E_n$) and eigenfunctions ($\Psi_n$) for the
local states from which the crystal ground and low lying excited states are
built. 
Even though these eigenfunctions are expanded over millions of determinants,
they are dominated by  only few of them, that are usually considered as the
implicit basis of the effective models ($t-J$ or Hubbard in the present
case). From the effective Hamiltonian theory (for instance such as derived in
the quasi-degenerate perturbation theory~\cite{Lindgren}) one can extract the
main concept necessary to get the effective Hamiltonian parameters from the ab
initio results. Indeed, the effective Hamiltonian on the fragment should
verify {\em at the best} the following set of equations
\begin{equation} \label{eq:heff}
 H_{\text{eff}} {\cal P}\Psi_n = E_n{\cal P}\Psi_n
\end{equation}
where $E_n$ and $\Psi_n$ are the ab initio energies and eigenstates, and ${\cal
  P}$ is the projection operator on the local configurations supporting
$H_{\text{eff}}$. For instance, in the present case, these configurations can
be the determinants supporting the local singlet and triplet state in the
$t-J$ or Hubbard model on the $a_{1g}$ orbitals. The $H_{\text{eff}}$
parameters are thus obtained from a mean-square fit minimization of
equation~\ref{eq:heff}.  These mean-square fit equations can sometime
be inversed and the parameters obtained from simple algebraic equations. This
is for instance the case for the effective exchange $J$ that is  the
singlet-triplet excitation energy of a $\rm Co_2 O_{10}$ fragment. Similarly,
the effective transfer can be simply expressed (see equation~\ref{eq:t} in
section~\ref{res:tJ}) from the same excitation energy and the knowledge of the
wave-functions.

One should however remember that the values of the effective integrals
strongly depend on the nature of the effective model. Indeed the processes
renormalizing the integrals in one model can be explicit in another and one
should take this into account in order to avoid double counting. Let us
illustrate our purpose on the $t_{2g}$ orbital energy splitting. In a Hubbard
model based on the cobalt $3d$ orbitals the $\rm CoO_6$ fragment low lying
states can be expressed as
\begin{eqnarray*}
\Psi_0&=& |e_{g1}^\prime \overline{e_{g1}^\prime} e_{g2}^\prime \overline{e_{g2}^\prime}
a_{1g} \rangle
\\ \text{with energy} && 
\varepsilon_{a_{1g}} + 2\varepsilon_{e_{g1}^\prime} +
2\varepsilon_{e_{g2}^\prime} + 2U + 8V - 4J_H \\[2ex]
\Psi_1&=&|e_{g1}^\prime  e_{g2}^\prime \overline{e_{g2}^\prime}
a_{1g} \overline{a_{1g}}\rangle
\\ \text{with energy} &&
2\varepsilon_{a_{1g}} + \varepsilon_{e_{g1}^\prime} +
2\varepsilon_{e_{g2}^\prime} + 2U + 8V - 4J_H\\[2ex]
\Psi_2&=&|e_{g1}^\prime \overline{e_{g1}^\prime} e_{g2}^\prime 
a_{1g} \overline{a_{1g}} \rangle
\\ \text{with energy} &&
2\varepsilon_{a_{1g}} + 2\varepsilon_{e_{g1}^\prime} +
\varepsilon_{e_{g2}^\prime} + 2U + 8V - 4J_H
\end{eqnarray*}
where $U$ is the repulsion of two electrons on the same orbital, $V$ and $J_H$
the repulsion and exchange of two electrons on different $3d$ orbitals. It
results that if $E_0$, $E_1$ and $E_2$ are the ab initio energies associated
with these states, the ab initio spectrum should be associated with the
orbital energies differences
$$E_1-E_0= \varepsilon_{a_{1g}} - \varepsilon_{e_{g1}^\prime} 
\quad \text{and}\quad 
E_2-E_0= \varepsilon_{a_{1g}} - \varepsilon_{e_{g2}^\prime} 
$$
On the contrary if one extends the model in order to explicitely treat the
differences ---~according to the nature of the $3d$ orbitals~--- in the
two-electrons two-orbitals repulsion and exchange integrals, then the ab
initio spectrum does no more correspond to pure orbital energy differences but
rather to
\begin{eqnarray*}
E_1-E_0&=& \varepsilon_{a_{1g}} - \varepsilon_{e_{g1}^\prime} 
+ 2 (V_{a_{1g} \, e_{g2}^\prime} -  V_{e_{g1}^\prime \, e_{g1}^\prime}) 
\\ && \qquad  \qquad 
- (J_{H\;a_{1g} \, e_{g2}^\prime} - J_{H\;e_{g1}^\prime \, e_{g1}^\prime} )
\\
E_2-E_0&=& \varepsilon_{a_{1g}} - \varepsilon_{e_{g2}^\prime} 
+ 2 (V_{a_{1g} \, e_{g1}^\prime} -  V_{e_{g1}^\prime \, e_{g1}^\prime} )
\\ && \qquad  \qquad 
- (J_{H\;a_{1g} \, e_{g1}^\prime} - J_{H\;e_{g1}^\prime \, e_{g1}^\prime} )
\end{eqnarray*}
In these equations the two electrons part is non zero as soon as the
$e_{g}^\prime$ orbitals are not pure $t_{2g}$ ones, but rather $t_{2g}$-$e_g$
hybrids, as it is the case as soon as the trigonal distortion sets in (see
reference~\onlinecite{nous2} and section~\ref{res:hyb} for the present
results). Whether the Racah's parameters are explicitely included in the model
or the spherical approximation  used, one should thus use different
$\varepsilon_{a_{1g}} - \varepsilon_{e_{gi}^\prime}$ effective orbital energy
differences in order for the effective Hamiltonian to fit the same ab initio
spectrum. In the case of $\rm CoO_2$ layers it means different ordering for
the $a_{1g}$ and $e_{g}^\prime$ orbitals (see
reference~\onlinecite{nous2}). As usually done, we will use the spherical
approximation in this work and renormalize the orbital energies by the effect
of the Racah's parameters.

\section{Results} \label{resultats}
\subsection{On site orbital energy splitting and charge order}
\label{res:orb}

A strong controversy has been going on in the literature about the cobalt 3d
orbital splitting. Indeed, authors did not agree on the relative order of the
$t_{2g}$ orbitals under the trigonal distortion of the regular octahedron
($T_{2g} \longrightarrow A_{1g}\oplus E_g$). The origin of the controversy was
the fact that the crystalline field theory~\cite{Mae} as well as some LDA
calculations~\cite{DFT2} found the $a_{1g}$ orbital of lower energy than the
$e_g^\prime$ ones, while other density functional calculations~\cite{DFT1}, as
well as quantum chemical calculations~\cite{nous1} or ARPES experimental
results~\cite{ARPES} found them of reverse order.  We recently
showed~\cite{nous2} that the origin of the controversy was an incorrect
treatment of the exchange and correlation integrals within the 3d
shell. Indeed, only when these effects are taken into account with their whole
complexity ---~that is when the orbital dependence of the two-electron
two-orbital on-site repulsion and exchange are correctly treated~---, the
relative order between the $a_{1g}$ and $e_g^\prime$ orbitals issued from the
$t_{2g}$ is correctly predicted. Unfortunately this is not the case in the LDA
approximation (essentially due to the local exchange) and thus LDA predicts
incorrect $a_{1g}$--$e_g^\prime$ ordering.  This splitting is however one of
the crucial parameter for the determination of the nature of Fermi level
states as recently shown by Marianetti {\it et al}~\cite{DMFT_mar}. Indeed,
they showed (using DMFT calculations) that the existence of the $e_g^\prime$
pockets in the Fermi surface of the $\rm Na_x CoO_2$ compounds are directly
related to the this splitting. One can thus conclude that the discrepancy
between LDA and ARPES Fermi surfaces can be directly related to the improper
treatment of the b Racah parameter in LDA. In fact, one can expect that an
orbital dependant LDA+U, where the whole complexity of the Racah parameters
would be included, should yield Fermi surfaces in agreement with the
experimental ones.

\begin{figure}[h]
\resizebox{8cm}{!}{\includegraphics{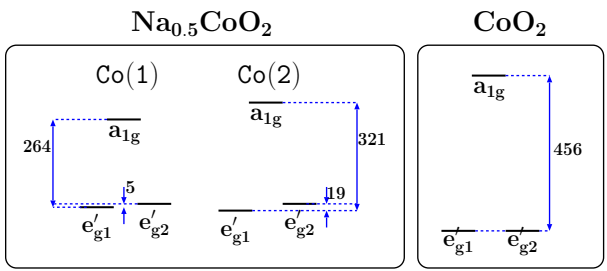}}
\caption{Schematic representation of the effective cobalt $t_{2g}$ orbital
  energies in meV. Let us recall that the Co(1) cobalt of the $x=0.5$ system
  is located directly under or on top a $\rm Na^+$ ion. }
\label{f:orbsplt}
\end{figure}

We thus computed the $t_{2g}$ orbitals splitting in the two compounds, for the
different cobalt sites. These values can be extracted from the spectroscopy of
the $\rm Co O_{6}$ embedded fragments.  Indeed, the excitation energy between
the three cobalt states, where one hole is located on one of the $t_{2g}$
orbitals, yield the effective splittings between the $a_{1g}$ and $e_g^\prime$
orbitals.  The relative order of the $t_{2g}$ orbitals is displayed on
figure~\ref{f:orbsplt}.
One sees immediately that the $a_{1g}$ orbitals are always higher than the
$e_g^\prime$ ones as can be expected from reference~\onlinecite{nous2}. In the
$x=0.5$ system, both sites exhibit $a_{1g}$ orbitals about 250--300~meV higher
than the $e_g^\prime$ ones. Similarly, in our dimer calculations we see that
the $\rm Co^{4+}(1)$ ion is globally of lower energy than the $\rm Co^{4+}(2)$
ion. Indeed, our calculations yield a global charge order of about 0.17
electrons in favor of the Co(1) site.  This result is in agreement with the
common sense expectations that will favor a smaller valence for the cobalt
with the largest number of $\rm Na^+$ cations neighbors. Indeed, computing the
electrostatic potential difference between the two cobalt sites one finds the
Co(1) site about 315~meV lower than the Co(2) site.  This result is however in
contradiction with the Bond Valence Sum (BVS) model, since the later yields a
charge ordering of 0.12 electron in favor of the Co(2) site~\cite{struct05},
however it is in agreement with the LDA+U calculations~\cite{pickett05} ($\sim
0.16 \bar e$ in favor of Co(1)). 
In the pure $\rm CoO_2$ system, the $a_{1g}$-$e_g^\prime$ orbital splitting is
very large, much larger than in any of the other $\rm Na_xCoO_2$ systems,
even larger than in the superconducting compound. This very large splitting
can be easily explained from (i) the large trigonal distortion (the largest
observed in this family) and (ii) the very small Co--O distances. Indeed, it
was shown in reference~\onlinecite{nous2} that (i) the orbital splitting
increases with increasing trigonal distortion and (ii) that the slope of the
splitting as a function of the distortion increases linearly with decreasing
Co--O distances.

\subsection{Orbital hybridization} \label{res:hyb}
Let us now focus on the cobalt 3d orbital hybridization.  These orbitals can
hybridize in two ways~: with the oxygen ligands 2p orbitals, but also within
the cobalt 3d shell between the $t_{2g}$ and $e_g$ sets of the regular
octahedron. Indeed, it was shown (see ref.~\onlinecite{nous2} for details)
that not only this hybridization is symmetry allowed by the trigonal
distortion of the octahedron, but also that it is responsible for the lowering
of the $e_{g}^\prime$ orbitals compared to the $a_{1g}$ one. The
$t_{2g}$--$e_g$ hybridization angle is reported in table~\ref{t:hybd} and
found to be non negligible, of the same order of magnitude as found in the
other $\rm CoO_2$-based systems. The $t_{2g}$--$e_g$ hybridization is larger
for $x=0$ than for $x=0.5$, as can be expected from the larger trigonal
distortion.
\begin{table}[h]
\begin{tabular}{c|ddd}
Orbital & \multicolumn{2}{c}{$\rm Na_{0.5}CoO_2$} & 
           \multicolumn{1}{c}{$\rm CoO_2$}  \\
& \multicolumn{1}{c}{Co(1)} & \multicolumn{1}{c}{Co(2)} & \\
\hline
$e_{g1}^\prime$ & 11.59 & 11.54 & 11.86 \\
$e_{g2}^\prime$ &  8.21 & 10.18 & 11.86 \\
\end{tabular}
\caption{ $t_{2g}$--$e_g$ mixing angle (degrees) in the occupied
$e_g^\prime$ cobalt orbitals.}
\label{t:hybd}
\end{table}

As far as the cobalt ligand hybridization is concerned, we found it to be weak
for the $a_{1g}$ and $e_g^\prime$ orbitals, with less than 10\% weight on the
oxygens. It is however quite large for the empty $e_g$ orbitals with 40 to
45\,\% weight on the neighboring oxygens.

\subsection{Interatomic interactions}\label{res:tJ}
Let us now focus on the interactions between two first neighbor cobalt
atoms. Many authors assume that the low energy physics of this compound is
determined by the $a_{1g}$ orbitals only. The large $a_{1g}$--$e_g^\prime$
orbital splitting found in our calculations seem to support this assumption.
We thus computed the effective integrals both for a $t-J$ model and a one-band
Hubbard model.

\begin{figure}[h]
\resizebox{8cm}{!}{\includegraphics{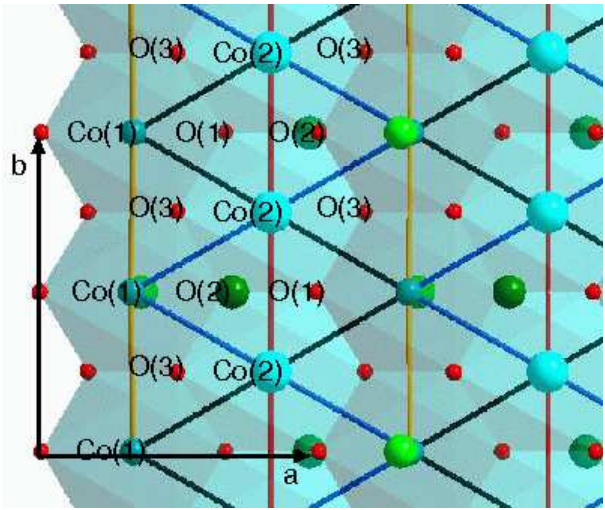}}
\caption{(color online) Schematic representation of the $\rm CoO_2$
  layers in the $\rm Na_{0.5}CoO_2$ compound. There are two different
  cobalt sites represented in light and darker turkoise color,
  three different oxygen ligands and the four different Co--Co bonds
  represented in orange [Co(1)--Co(1)], red [Co(2)--Co(2)], dark blue
  [Co(1)--Co(2) bridged by O(2) and O(3) oxygens] and black
  [Co(1)--Co(2) bridged by O(1) and O(3) oxygens].}
\label{f:ab}
\end{figure}

Analyzing the low temperature crystallographic structure of $\rm
Na_{0.5}CoO_2$ (10~K data of reference~\onlinecite{struct05}), one sees that
there are four types of independent $\rm Co$--$\rm Co$ bonds (see
figure~\ref{f:ab}), the Co(1)--Co(1) bond, the Co(2)--Co(2) bond and two kinds
of Co(1)--Co(2) bonds, namely those where the cobalt--cobalt interactions are
mediated by one O(1) and one O(3) oxygen ligands and those where the
cobalt--cobalt interactions are mediated by one O(2) and one O(3) oxygen
ligands (see figure~\ref{f:ab}).
The effective magnetic  exchange integrals are   
\begin{eqnarray*}
J(11)=  -11\,\text{meV} & \quad J(22)=-27\,\text{meV} \\ 
J(12^{[13]})= -19\,\text{meV}  & J(12^{[23]})= -19\,\text{meV}
\end{eqnarray*}
where $J(ij)$ refers to the Co(i)--Co(j) bond and the $[ab]$ superscript
refers to the bridging oxygens.  For the pure $\rm CoO_2$ system the magnetic
coupling was computed to be much larger with 
$$ J=-52\,\text{meV}$$
In our notations, the negative integrals correspond to antiferromagnetic
coupling.  We thus find all exchange integrals, both in the $x=0$ and $x=0.5$
compounds, antiferromagnetic, in agreement with the neutron
scattering~\cite{neutron06} and NMR~\cite{CoO2_1,CoO2_3} data. One should
notice that the exchange integrals are strongly modulated according to the
local environment. Indeed, they differ by a factor larger than two for the
different bond of the $x=0.5$ system. It results that this system will be
described by a strongly inhomogeneous $t-J$ model. In the $x=0$ compound the
AFM exchange is very large, in agreement with NMR experimental
results~\cite{CoO2_1}.  The difference in magnitude between the
antiferromagnetic coupling of the $x=0$, $x=0.5$ and superconducting
system~\cite{nous1} can easily be explained by the variations in the Co--O
distances. Indeed, these distances control for a large part the super-exchange
term of the coupling that increases with decreasing distances. It results that
following De Vaulx {\it et al}~\cite{CoO2_1} we found that AFM fluctuations
increase with decreasing $x$.

\begin{figure}[h]
\resizebox{3.5cm}{!}{\includegraphics{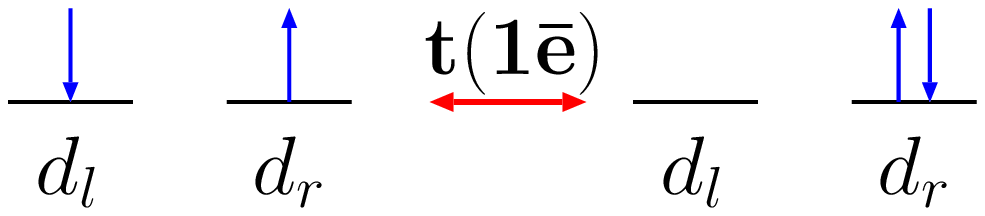}} \hspace*{1cm}
\resizebox{3.5cm}{!}{\includegraphics{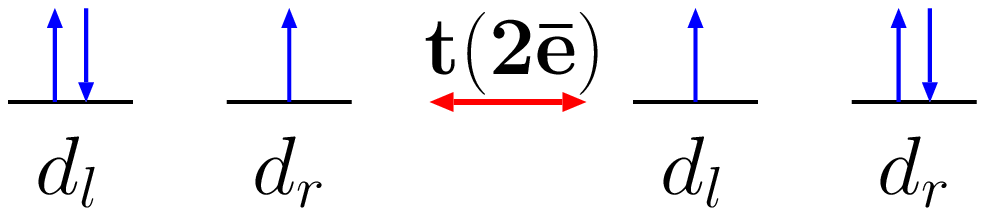}}
\caption{Schematic representation of the effective hopping integrals as
a function of the spectator electrons on the bond. $l$ and $r$ labels
refers to the arbitrary {\em ``left''} and {\em ``right''} atoms in
the fragment.}
\label{f:hop}
\end{figure}

Let us now focus on the hopping integrals. One can get them either from the
charge fluctuation in the singlet state of the $\rm Co^{4+}$--$\,\rm Co^{4+}$
fragments, or from the spectroscopy of the $\rm Co^{3+}$--$\,\rm Co^{4+}$
fragments. A simple analysis shows that the two hopping integrals are
somewhat different. In the first case, there is only one spectator
electron on the bond, while in the second case, there are two of
them (see figure~\ref{f:hop}). 
It results in a different evaluation of the hopping integrals according
to the number of spectator electrons.
\begin{eqnarray*}
t(1\bar e) &=& \langle d_l | h | d_r\rangle + 
               \langle d_l \bar d_l | 1/r_{12} |   d_r \bar d_l \rangle \\
t(2\bar e) &=& \langle d_l | h | d_r\rangle + 
               \langle d_l \bar d_l | 1/r_{12} |   d_r \bar d_l \rangle +
               \langle d_l \bar d_r | 1/r_{12} |   d_r \bar d_r \rangle
\end{eqnarray*}
where the $l$ and $r$ labels refers to the arbitrary {\em ``left''}
and {\em ``right''} atoms in the fragment, and $h$ is the single
electron Hamiltonian part.

$t(1\bar e)$ can be extracted from the $\rm Co^{4+}$--$\,\rm Co ^{4+}$
fragment. Indeed, mapping a single band Hubbard model on the computed
low-energy singlet and triplet states wave functions and energies, one
gets after a little algebra
\begin{eqnarray}
t(1\bar e)&=& J \frac{1}{2 \cos{\varphi} \tan{\alpha}} \label{eq:t}\\
\overline{U} -V &=& J \left( 1- \frac{1}{\cos{2\varphi} \tan^2{\alpha}} \right) \label{eq:U}
%
\label{eq:delta}
\end{eqnarray}
assuming the following form for the computed singlet and triplet wave
functions
\begin{eqnarray*} 
\psi_{Sg}&=& \cos{\alpha} 
\frac{|d_l\bar d_r\rangle + |d_r\bar d_l\rangle  }{\sqrt{2}} \\ &+&
 \sin{\alpha} \left( \cos{\varphi} 
\frac{|d_l\bar d_l\rangle + |d_r\bar d_r\rangle  }{\sqrt{2}} +
\sin{\varphi}
\frac{|d_l\bar d_l\rangle - |d_r\bar d_r\rangle  }{\sqrt{2}}
\right) \\ &+&
\text{small terms} \\[2eM]
\psi_{Tp}&=& \frac{|d_l\bar d_r\rangle - |d_r\bar d_l\rangle  }{\sqrt{2}}  
+
\text{small terms} 
\end{eqnarray*}
$J= E_{Sg} - E_{Tp}$. $\overline{U}-V$ is the effective on-site
repulsion of a simple Hubbard model. Let us remind that it
accounts for both the average on-site repulsion between the two sites,
$\overline{U}=\left( U_l+U_r\right) /2$, and the nearest neighbor
effective repulsion between the cobalt atoms, $V=V_{lr}$.
The (electron-)hopping integrals between the $a_{1g}$ orbitals thus come as 
\begin{eqnarray*}
t^{1\bar e}_{00}(11) = 84\,\text{meV}  & \quad
t^{1\bar e}_{00}(22) = 139\,\text{meV} \\
t^{1\bar e}_{00}(12^{[13]}) = 115\,\text{meV}&
t^{1\bar e}_{00}(12^{[23]}) = 111\,\text{meV}&
\end{eqnarray*}
for the $x=0.5$ compound and as
\begin{eqnarray*} t_{00}^{1\bar e} =  212\,\text{meV} \end{eqnarray*}
for the $x=0$ one.

Once again, one sees that the effective transfer integrals are strongly
dependant on the local environment of the cobalts and that the $\rm
Na_{0.5}CoO_2$ system is more inhomogeneous than first thought.  In the pure
compound, the effective transfer between the $a_{1g}$ orbitals of NN atoms is
larger than both in the $x=0.5$ and the superconducting system. The relative
variations of the Co--O and large Co--Co distances between these three systems
support the idea that the effective transfer integrals are dominated by the
ligand-mediated term rather than the direct hopping term, despite the short
metal-metal distances in these edge-sharing compounds.

The transfer integrals with two spectator electrons on the bond,
$t^{2\bar e}$, can be extracted from the low-energy states of the $\rm
Co^{3+}$--$\,\rm Co^{4+}$ fragments. There are six low energy states
according to the localization of the hole on the $a_{1g}$ and
$e_g^\prime$ orbitals. The associated effective Hamiltonian can be
written as
\begin{eqnarray}
 \label{eq:hd}
H &=&
\bordermatrix{\vspace*{1ex}
& a_{1g}^l & e_{g1}^{\prime~l} &e_{g2}^{\prime~l} &&a_{1g}^{r} &
  e_{g1}^{\prime~r}& e_{g2}^{\prime~r} \cr 
& \varepsilon^l_0 & tp^l_{10} & tp^l_{20} && 
t^{2\bar e}_{00} & t^{2\bar e}_{10} & t^{2\bar e}_{20} \vspace*{1ex} \cr
& tp^l_{10} &\varepsilon^l_1 & tp^l_{12}  && 
  t^{2\bar e}_{10}  & t^{2\bar e}_{11} & t^{2\bar e}_{12} \vspace*{1ex} \cr
& tp^l_{20} & tp^l_{12} & \varepsilon^l_2 &&  
  t^{2\bar e}_{20} & t^{2\bar e}_{12} & t^{2\bar e}_{22} \vspace*{2ex}\cr
& t^{2\bar e}_{00} & t^{2\bar e}_{10} & t^{2\bar e}_{20} && 
  \varepsilon^r_0 & tp^r_{10} & tp^r_{20} \vspace*{1ex}\cr
& t^{2\bar e}_{10}  & t^{2\bar e}_{11} & t^{2\bar e}_{12} &&  
  tp^r_{10} &\varepsilon^r_1 & tp^r_{12} \vspace*{1ex} \cr
&  t^{2\bar e}_{20} & t^{2\bar e}_{12} & t^{2\bar e}_{22} && 
   tp^r_{20} & tp^r_{12} & \varepsilon^r_2 \cr
}       
\label{eq:H}
\end{eqnarray}
where $\varepsilon_i$ are the atomic effective orbital energies (see
figure~\ref{f:orbsplt}), $t^{2\bar e}_{ij}$ are the effective transfer
integrals (direct plus mediated by the oxygen ligands) between the
$3d_i$ orbital of one cobalt and $3d_j$ of the other. $tp_{ij}$ are
effective intra-atomic transfer integrals between the $d_i$ and $d_j$
orbitals of the same atom. One may be surprised to find such terms,
since the direct integrals are zero due to the $Y_l^m$ symmetry, however the
coupling with the bridging oxygen $2p$ orbitals yield in second order
perturbation theory an effective intra-atomic transfer term of 
\begin{eqnarray*}
tp_{ij} &=& - \sum_p {t_{i\,p}t_{j\,p} \over \Delta_{p}} 
\end{eqnarray*}
where $i$ and $j$ refers to the $3d$ orbitals of the same cobalt atom,
the sum over $p$ runs over all the ligand bridging orbitals,
$t_{i\,p}$ is the cobalt $3d_i$--ligand hopping integrals and
$\Delta_{p}$ is the excitation energy toward the ligand-to-metal
charge transfer configuration. For a more detailed description of the
underlying mechanism, one can refer to reference~\onlinecite{nous1}.
Summing the $tp_{ij}$ contributions coming from the six oxygens around
a cobalt atom, one can show that they exactly cancel out in a symmetric
system. In the present system, where the atomic $S_6$ symmetry is not
exactly respected, these terms will certainly not exactly cancel out,
however, one can expect that their sum will remain very weak.

Table~\ref{t:t2e} displays the different effective integrals involved
in matrix~\ref{eq:H} for the $x=0.5$ system. 
\begin{table}[h]
\begin{tabular}{c|llllll}
& \multicolumn{6}{c}{Inter-atomic transfers (meV)}\\
Bond         & $t^{2\bar e}_{00}$ & $t^{2\bar e}_{11}$ & $t^{2\bar e}_{22}$ & 
$t^{2\bar e}_{10}$ &  $t^{2\bar e}_{20}$ & $t^{2\bar e}_{12}$  \\ \hline
Co(1)--Co(1) & 225  &   -261 &   -4  & -22  & -18 &  -27 \\
Co(2)--Co(2) & 281  &   -325 &   -11 & $\pm$ 10  & -18 &  $\pm$ 10 
\end{tabular} 
\caption{Effective (electron-)hopping integrals between $a_{1g}$ and
$e_g^\prime$ orbitals of two neighboring cobalt atoms (in meV).}
\label{t:t2e}
\end{table}
One sees immediately that the dominant transfer integrals are the
inter-atomic hopping between two $a_{1g}$ orbitals and two
$e_{g1}^\prime$ orbitals, in agreement with what we found
on the superconducting compound. All other integrals are relatively
weak ($<30\rm meV$) and can probably be omitted in a simple model.
Comparing these values with what we found on the superconductor
system, one sees that they are of the same order of magnitude, except
for the inter-atomic hopping between the $a_{1g}$ orbital of one
cobalt and the $e_{g2}^\prime$ orbital of the other. Indeed, the
larger value found in the superconducting systems (-53~meV) is now
spread over the three $t^{2\bar e}_{10}$, $t^{2\bar e}_{20}$ and
$t^{2\bar e}_{12}$ hopping terms, due to local symmetry breaking. 

\subsection{On-site repulsion}\label{res:U}

Let us now examine what we find for on-site the repulsion $U$ from our
calculations. Indeed, as detailed in equation~\ref{eq:U}, the value of
$\overline{U}-V$ can be extracted from the singlet and triplet energies and
wave functions on the $\rm Co^{4+}$--$\,\rm Co^{4+}$ fragments. Results for a
single band model are presented in table~\ref{t:U}.
\begin{table}[h!]
$$  \begin{array}{c|cc}
      \text{Compound} & U-V & (U-V)/t \\[1ex]
      \hline
      \rm CoO_2 & 3.71 & 15.6 \\[1ex]
      \rm [Na_{0.35}CoO_2][H_2O]_{1.3}~\cite{ThSylvain}&  3.65 & 19.4 \\[1ex]
      \rm Na_{0.5}CoO_2 &&\\
      \hfill U(1)-V(11) &  2.61 & 31.1 \\
      \hfill U(2)-V(22) &  2.86 & 20.6 \\ 
      \hfill \overline{U} - V(12^{[13]}) &  2.76 & 24.0 \\
      \hfill \overline{U} - V(12^{[23]}) &  2.56 & 23.1 
    \end{array}
$$
    \caption{\it Effective on-site Hubbard repulsion (eV) and correlation strength
      for different $\rm Na_xCoO_2$ compounds. $\overline{U} = [U(1)+U(2)]/2 $}
    \label{t:U}
\end{table}
One sees immediately that the Hubbard on-site repulsion decreases with
increasing sodium content, that is with increasing number of electrons in the
$\rm CoO_2$ layers. Once again correlating this fact with the Co--O distances
one sees that the larger the Co--O distances, the weaker the cobalt on-site
repulsion. This can be understood by the fact that a larger volume of the 
coordination shell around the metal atom will allow its magnetic orbitals to
spread over some more and this reduce the intra-orbital repulsion. 
If one now looks at the correlation strength ($U/t$) rather than the on-site
repulsion value, one sees that the tendency is reverse. Indeed, the larger the
sodium content, the larger the correlation strength. It is easy to see that
the opposite variation of the  correlation strength and the one-site
repulsion, as a function of $x$, is again due to the Co--O distances. Indeed, if
an increase of the metal coordination sphere induce a decrease of the
effective $U$, it induces a much stronger reduction of the effective hopping
integral $t$, thus resulting in weaker electronic correlation. This results
fit quite well with what was originally supposed (from experimental
observations) in these systems, that is a  decrease of the correlation strength
with $x$. However it seems hard to say that these systems are weakly
correlated, since the smallest $U/t$ ratio is 15.

\section{Conclusion}\label{conclusion}

We determined the effective on-site and nearest-neighbor coupling parameters
for the the $\rm Na_{0.5} Co O_2$ and $\rm CoO_2$ compounds within Hubbard and
$t-J$ models.  The effective integrals and orbital energies were computed
using ab-initio quantum chemical methods treating exactly the strong
correlation effects within the cobalt $3d$ shell and the screening effects up
to the double-excitations.

The Hubbard on-site repulsion and the correlation strength $U/t$ were found to
vary in opposite ways with the sodium content. Indeed, when doping the $\rm
CoO_2$ layer, the $t$ value decreases more rapidly than the $U$ value and the
resulting correlation strength increases. For the $x=0$ limit, however, the
system is still strongly correlated with $U/t\sim 15$. This result is in
disagreement with Onoda's conclusion that the $\rm CoO_2$ compound is a weakly
correlated metal~\cite{CoO2_2,CoO2_3}. On the contrary, Julien~\cite{CoO2_1}
concludes the system is a strongly correlated metal close to the Mott phase
transition. Our results clearly do agree with the latter hypothesis, indeed,
it was shown using DMFT calculation that the Hubbard model reach the Mott
transition in a triangular lattice at a critical correlation strength of
$U_c/t_c=15$~\cite{triang_dmft}. Our evaluation of the correlation strength~:
$15.6$, clearly situates the $\rm CoO_2$ compound very close to the Mott
transition. Similarly, our results are in agreement with NMR
measurements~\cite{CoO2_1,CoO2_3} that sees an increase of the
antiferromagnetic coupling with decreasing sodium content. However,
contrarily to Kawasaki {\it et al}~\cite{CoO2_3} we do not find the pure $\rm
CoO_2$ compound behaving on a specific way. Indeed, we found it to display the
largest antiferromagnetic coupling among all the systems we studied.

We determined the ligand field splitting for the different
crystallographically independent cobalts of the $\rm Na_{0.5} Co O_2$ and $\rm
CoO_2$ compounds.  As for the superconducting compound, we find that the
$a_{1g}$ orbitals are of higher energy than the $e_{g}^\prime$ ones, yielding
a ${e_g^\prime }^4\,{a_{1g}}^1$ cobalt atomic configuration. Comparing the
$t_{2g}$ splitting in these two compounds with the splitting in the
superconducting compound, we see that it follows the behavior predicted in our
previous work~\onlinecite{nous2} both as a function of the amplitude of the
trigonal distortion and as the size of the cobalt first coordination shell.
Indeed, in the $\rm CoO_2$ compound, the average distance between the oxygen
and cobalt planes is 0.85\AA{} while it is 0.88\AA{} in the superconducting
system and ranges between 0.95 and 0.98\AA{} in the $x=0.5$ compound. As this
distance increases with increasing electron doping of the $\rm CoO_2$ layer,
the splitting between the $e_{g}^\prime$ and $a_{1g}$ orbitals decreases,
being maximum for the pure system. Comparing this orbital order with the LDA
or LDA+U calculations one sees that it is reverse order. Indeed, tight-binding
fitting of the LDA+U calculations yield a higher hole energy by about 0.01eV
when located in the $a_{1g}$ orbital~\cite{fit}. Why do LDA calculations yield
a reverse order between the $a_{1g}$ and $e_g^\prime$ atomic orbitals than
wave-function cluster calculations? The answer is in the treatment of the
Coulomb and exchange terms within the 3d shell. Indeed, in LDA calculations
the correlation and exchange contributions are treated using a local
approximation. It means that the small differences in energy that should occur
in these term according to the nature of the occupied 3d orbitals (the $b$
Racah's parameter) is ignored. This is even worse with LDA+U calculations
since the Hartree term is also replaced and a unique $U$ (or rather $U$, $V$
and $J_H$) is used while the Coulomb and exchange integrals between two
different $3d$ orbitals depend on the nature of the considered orbitals. These
integrals differences are quite small and, in most cases, the averaging is
valid enough.  In the present case, however, it can be shown that the
2-electron contribution to the orbital energy that is responsible for the
$a_{1g}$--$e_g^\prime$ inversion in reference~\onlinecite{nous2}, is a direct
function of the Racah's $b$ integral, and of the Coulomb and exchange
integrals differences. The effects of these imperfect exchange and correlation
treatment within the 3d shell, and thus of the resulting orbital order
between the $a_{1g}$ and $e_g^\prime$ orbitals is to pop up the band issued
from the later in comparison with the bands issued from the former. Indeed, it
was shown by Marianetti {\it et al}~\cite{DMFT_mar} (using DMFT calculations)
that when the $a_{1g}$ orbital is supposed of higher energy (lower hole
energy) than the $e_g^\prime$ orbital, the $e_g^\prime$ pockets in the Fermi
surface disappear.  More recently Bourgeois {\it et al}~\cite{DMFT2}
showed that using orbital splitting of the same sign and order of magnitude as in the
present work, DMFT Fermi surfaces and band structure nicely fit the ARPES data
(they studied different systems ranging from $x=0.3$ to $x=0.7$). 
In view of our results and the above analysis we think
that the $e_g^\prime$ Fermi pockets found in LDA calculations and not observed
in ARPES experiments are due to an intrinsic feature of the LDA exchange and
correlation functional and thus that these pockets are true artefacts of the
method.

Finally, one should remember the strong variation of the effective exchange
and hopping integrals as a function of the local distortions. This fact can be
of importance in the understanding of the changes in magnetic behavior, as a
function of $x$, in the $\rm Na_xCoO_2$ family. It would thus be of interest
to see if the local distortions observed for larger $x$ values will be able to
reverse the sign of the magnetic exchange as suggested by neutron diffraction
experiments.

\acknowledgments The authors thank Daniel Maynau for providing us with
the CASDI suite of programs. These calculations where done using the
CNRS IDRIS computational facilities under project n$^\circ$1842 and
the CRIHAN computational facilities under project n$^\circ$2007013.


\end{document}